\documentclass{appolb}
\usepackage{graphicx}

\begin{document}

\title{The AFP Project\thanks{Presented at the Cracow Epiphany Conference on the first year of the LHC,\\Cracow, Poland, January 10--12, 2011}}
\author{R. Staszewski\thanks{On behalf of the ATLAS Collaboration}
\address{Institute of Nuclear Physics Polish Academy of Sciences\\
ul. Radzikowskiego 152, 31-342 Krak\'ow, Poland.}
}

\graphicspath{{diagrams/}}

\maketitle \begin{abstract}
AFP is a project to extend the diffractive physics programme of the ATLAS
experiment by installing new detectors that will be able to tag forward protons
scattered at very small angles. This will allow us to study Single Diffraction,
Double Pomeron Exchange, Central Exclusive Production and photon-photon
processes. This note presents the physics case for the AFP project and briefly
describes the proposed detector system.
\end{abstract}


\section{Introduction}

\begin{figure}[h]
  \centering
  \includegraphics[width=0.49\textwidth]{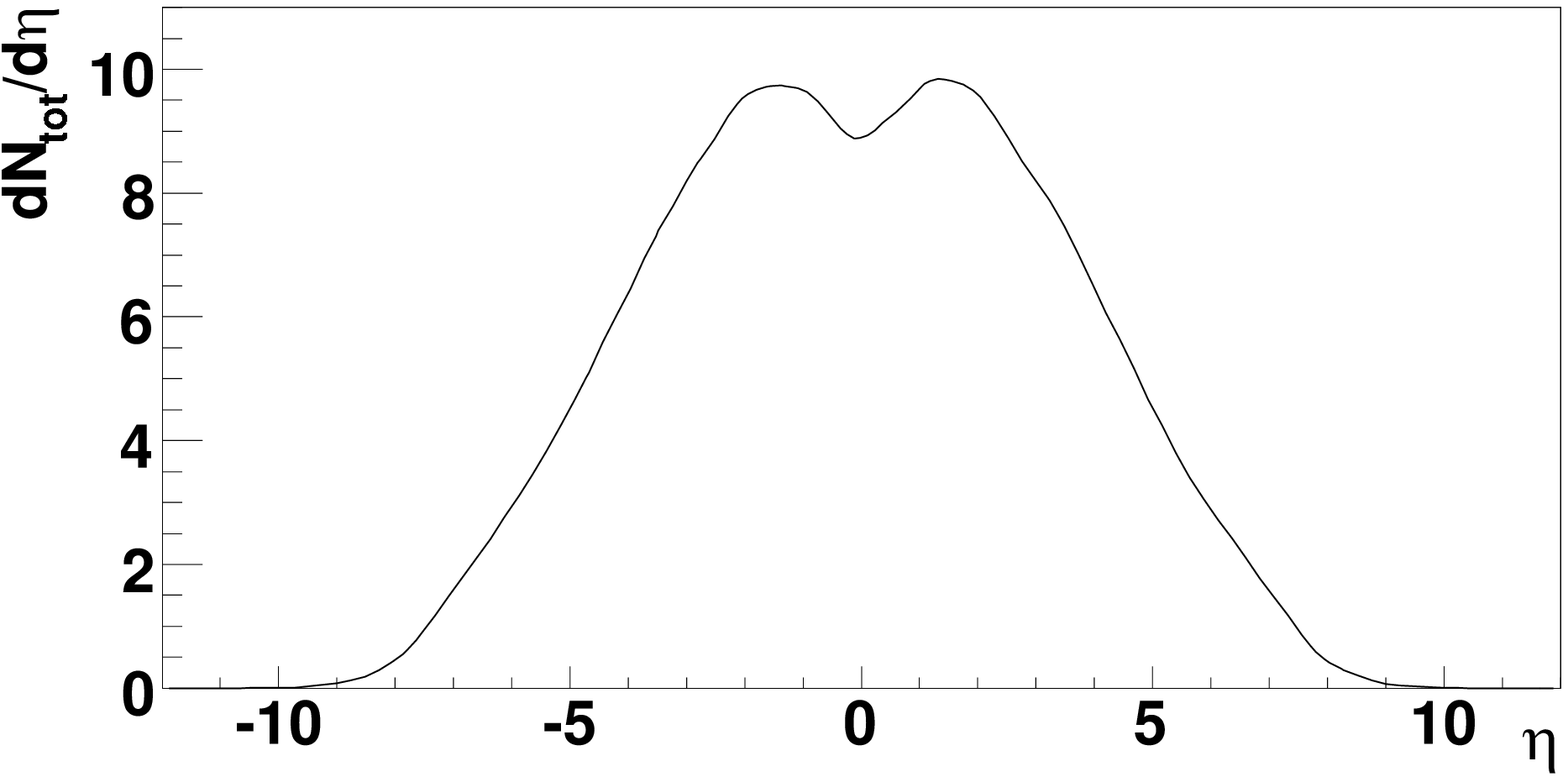}
  \hfill
  \includegraphics[width=0.49\textwidth]{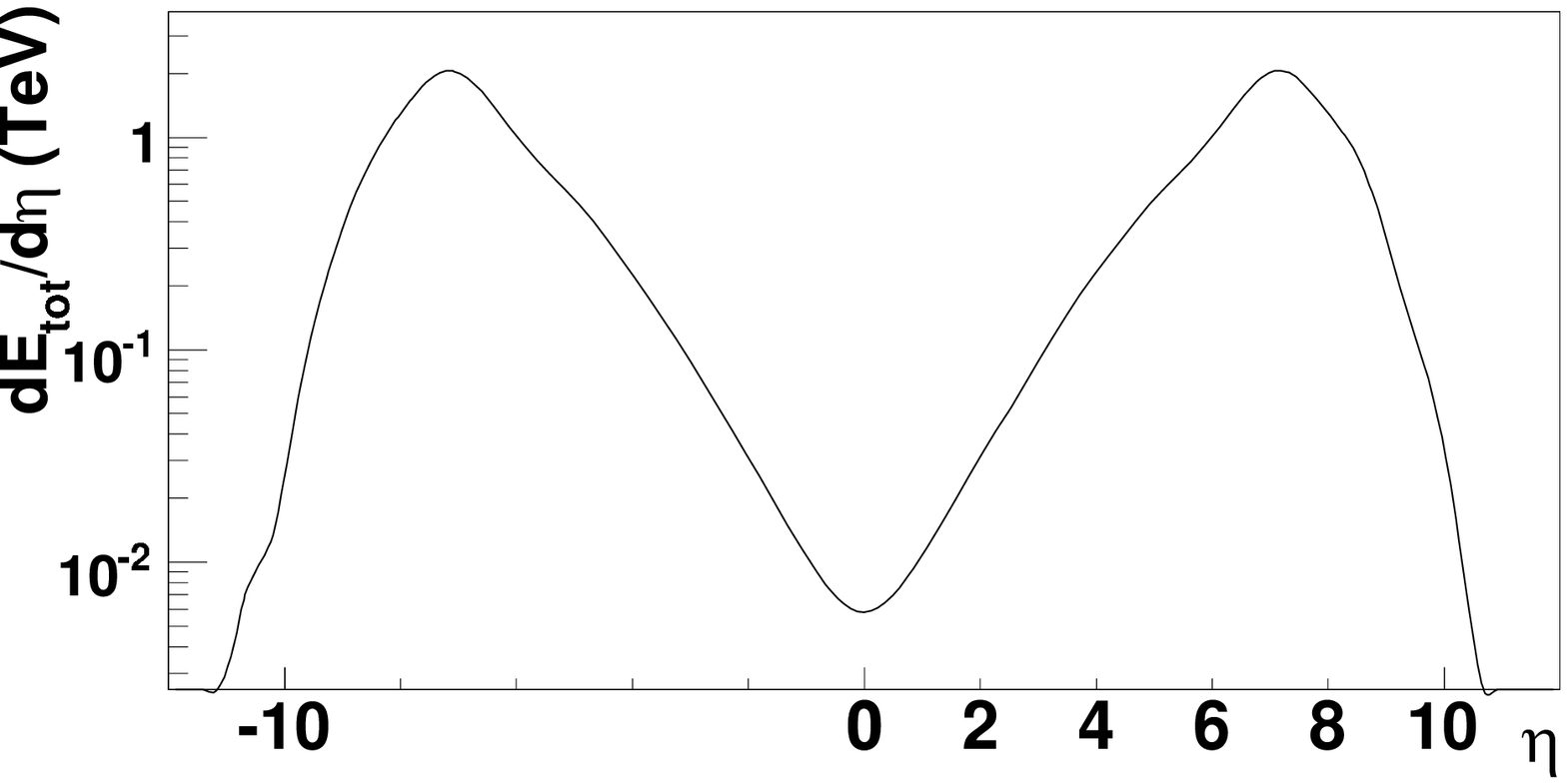}
  \caption{Multiplicity (left) and energy (right) distributions (adapted from \cite{d'Enterria:2007dt}).}
  \label{fig:NEdist}
\end{figure}

In the high energy $pp$ collisions at the LHC most attention is usually
paid to the central rapidity region, \textit{i.e.} where the most of the
particles are produced and where the most of the high $p_T$ signal of
new physics is expected. However, the energy emitted in such
interactions is distributed mainly in the forward direction and usually most of it
 disappears in the accelerator beampipe (see Fig. \ref{fig:NEdist}).

Forward physics is devoted to studies of high rapidity regions and includes
many interesting topics like elastic scattering, diffraction, low-x QCD and
Central Exclusive Production, photon-photon interaction, the two last being the
main motivation for the AFP project. 

\section{Central Exclusive Production}

Central Exclusive Production (CEP) is a very interesting class of processes in
which the two interacting protons are not destroyed during the interaction but
survive into the final state (intact protons, forward protons). Such situation
takes place for example during elastic scattering, but can happen also in
quasi-elastic processes when, apart from the two protons, there is an
additional particle (or particles) in the final state.

\begin{figure}[h] \hfill \includegraphics[width=0.3\textwidth]{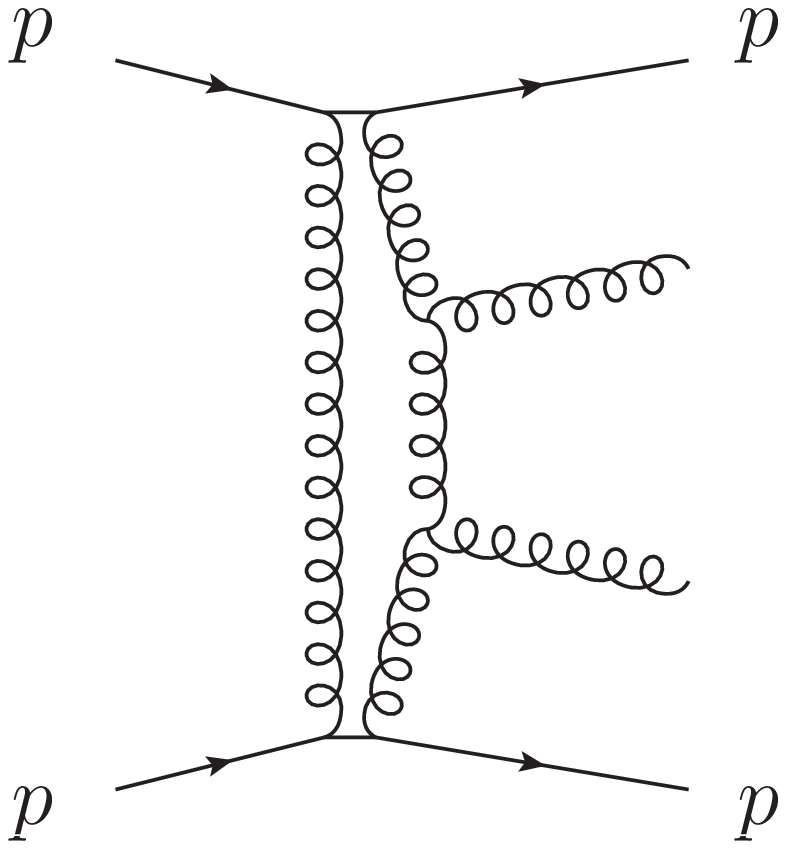}
  \hfill \includegraphics[width=0.3\textwidth]{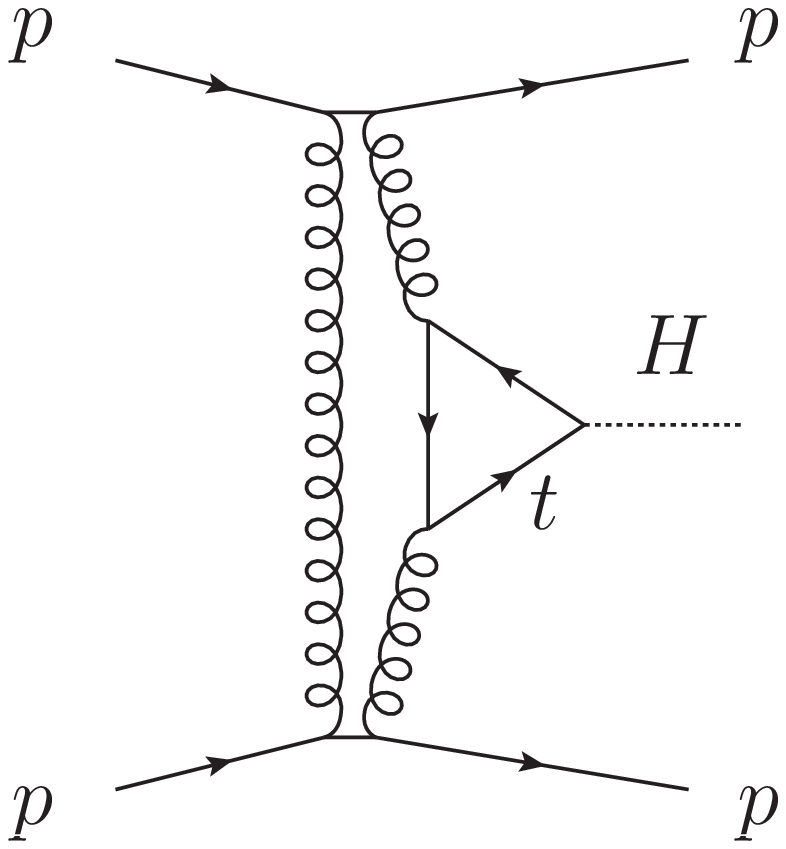} \hfill\hfill
  \caption{Feynman diagrams of exclusive jets (left) and Higgs (right)
  production.} \label{fig:jjHdiag} \end{figure}

This is a very rare situation and can take place only when the protons interact
coherently via an emission of a color singlet object. Such interaction can be
of an electromagnetic or a strong nature, occurring via a photon or a Pomeron
exchange, respectively. In hard processes the Pomeron exchange is usually
modelled, at the lowest order, by an exchange of a colorless two-gluon system.
Feynman diagrams for two important examples are presented in Fig.
\ref{fig:jjHdiag} -- exclusive production of jets (left) and the Higgs boson
(right). One can notice a similar structure: two gluons are emitted from each
proton, one of the gluons is involved in the hard subprocess ($gg\rightarrow
gg$ or $gg\rightarrow H$), the second one screens the process.  Since no color
is exchanged between the protons and the central state (jets or Higgs), it is
possible that the protons stay intact after the interaction. It is important to
point out that in the Central Exclusive Production, contrary to the Double
Pomeron Exchange processes, there are no Pomeron remnants in the final state.

The exclusive jet production has already been studied at the Tevatron
\cite{Aaltonen:2007hs} and there is a hope for observing the exclusive Higgs at
the LHC. This would have several advantages over the ``standard'' searches.
First of all, the exclusive $b\bar b$ jets production (background to
exclusive $H\rightarrow b\bar b$) is suppressed by the $J_z=0$ selection rule
\cite{Khoze:2000cy}. Secondly, for such a process it is possible to detect all
particles in the final state, provided the forward protons are detected. In
such a way the kinematics of the event is fully constrained, which leads to a
good resolution for the Higgs mass measurement in a wide range of masses
\cite{Staszewski:2009sw}.  Thirdly, observing the Higgs in the exclusive
production mode indicates strongly that it is a $0^{++}$ particle, since the
amplitudes for other spin states are much smaller. The drawback of this
approach for the Standard Model (SM) Higgs is a very small production cross
section, which is predicted (with very large uncertainty
\cite{Dechambre:2011py}) to be of the order of a femtobarn.  However, in
supersymetry (SUSY) scenarios the cross section can be much bigger.

\begin{figure}[h]
  \hfill
  \includegraphics[width=0.3\textwidth]{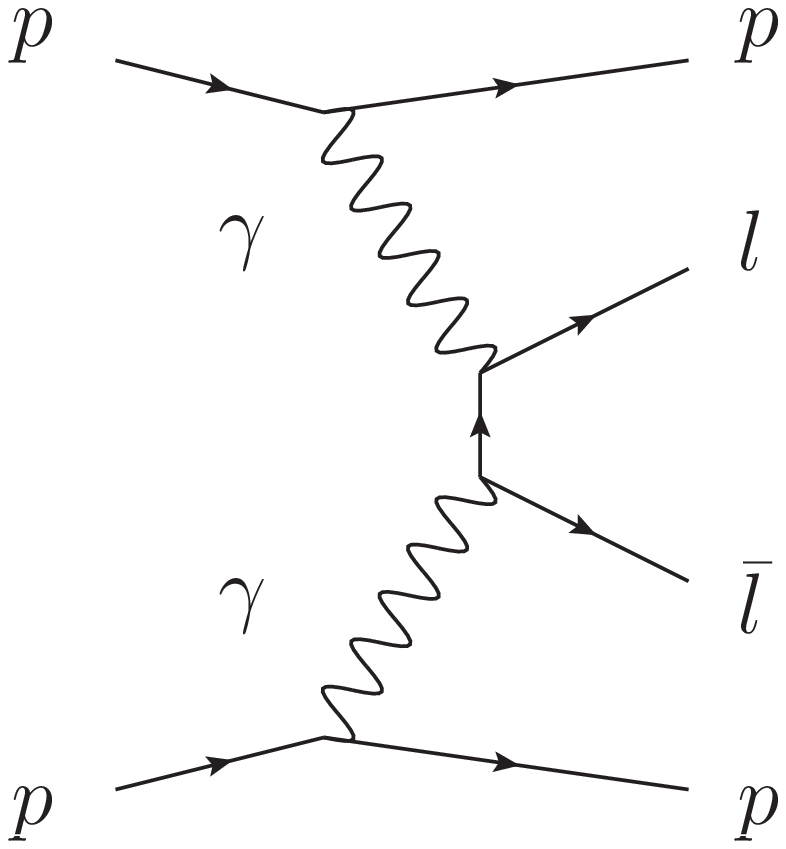}
  \hfill
  \includegraphics[width=0.3\textwidth]{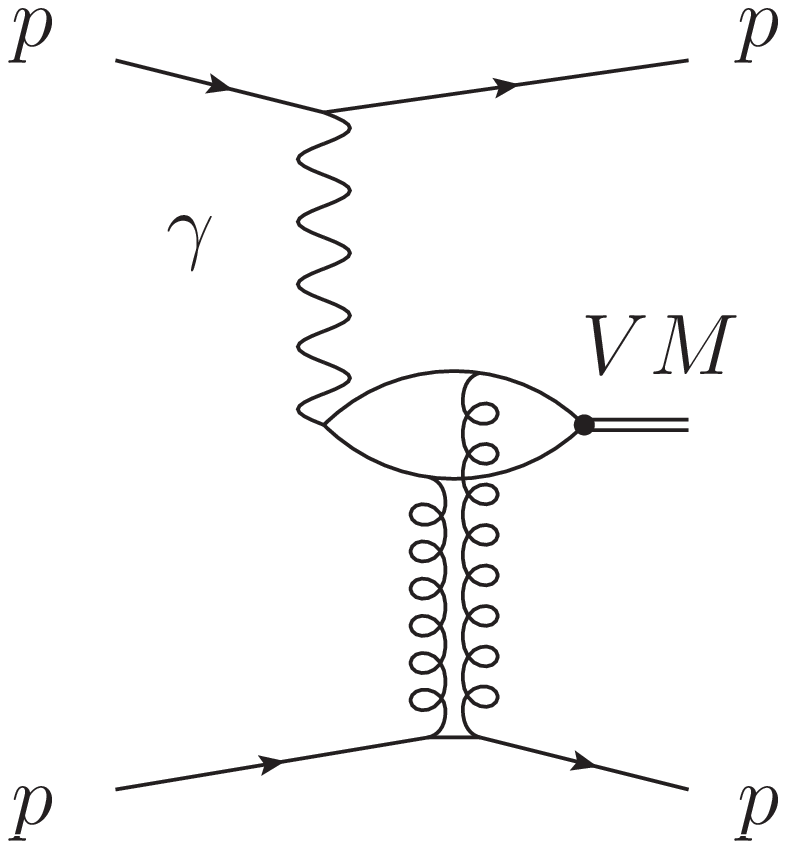}
  \hfill\hfill
  \caption{Feynman diagrams of diphoton exchange (left) and exclusive
  photoproduction (right) processes.}
  \label{fig:diphot_photo}
\end{figure}

As mentioned before the exclusive production can be not only a strong, but also
an electromagnetic interaction.  A Feynman diagram for an example of such a
process is shown in Fig. \ref{fig:diphot_photo} (left). One can see the photons
that are emitted from the protons, interact with each other and produce a
lepton pair. This kind of process is called a photon-photon interaction and the
same mechanism can lead to production of pairs of $W$ bosons, sparticles,
\textit{etc}. Since these are the QED processes, they are well understood and
the prediction for cross sections have very small uncertainty. Therefore, they
can be used for a precise luminosity determination at the LHC
\cite{Krasny:2006xg}.  Also, similarly to the exclusive Higgs case, the
detection of the outgoing protons can give a precise determination of the central
system mass, which can be used for SUSY studies \cite{Schul:2008sr}.

There exists also a third possibility of the exclusive process. This is the
exclusive photoproduction and it combines the electromagnetic and strong
interactions. Here, one of the protons emits a photon that interacts with the
Pomeron emitted by the second proton, see Fig. \ref{fig:diphot_photo} (right).
The result is a vector meson and two intact protons in the final state.

\section{Anomalous Gauge Bosons Couplings}

\begin{figure}[h]
  \includegraphics[width=0.3\textwidth]{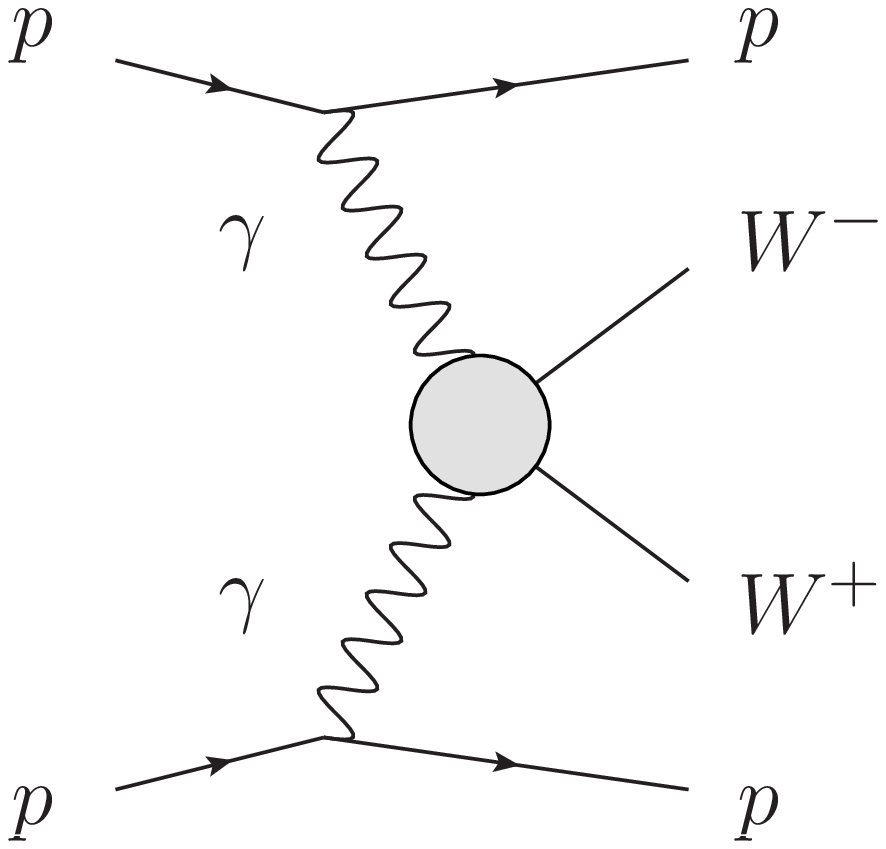}
  \hfill
  \includegraphics[width=0.3\textwidth]{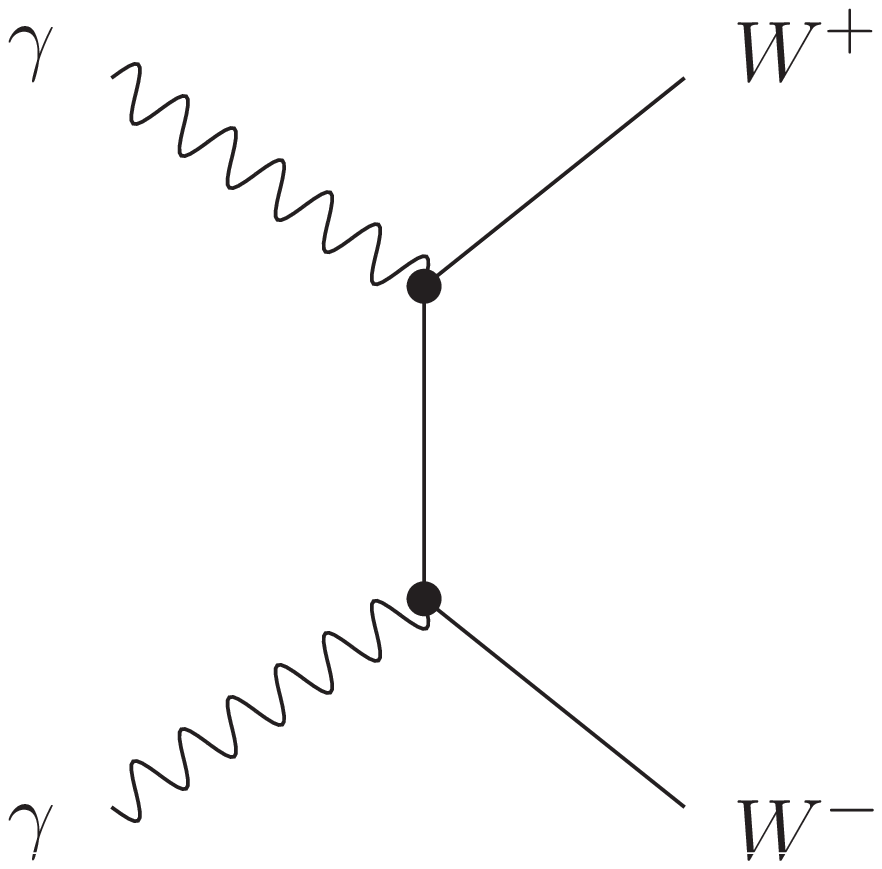}
  \hfill
  \includegraphics[width=0.3\textwidth]{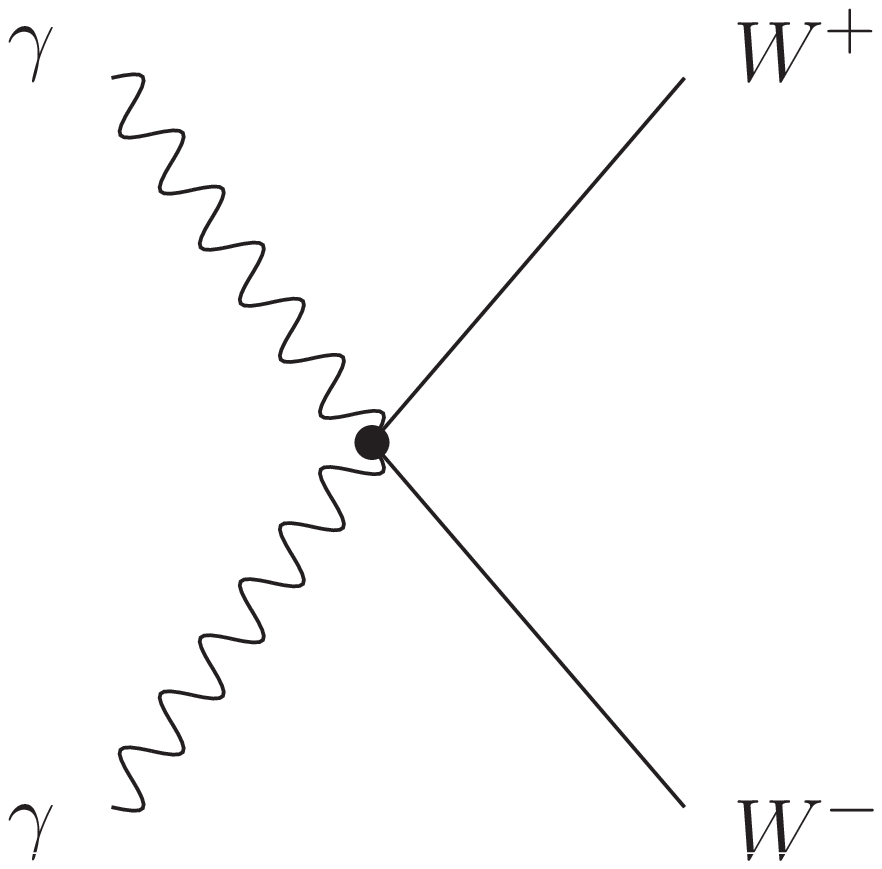}
  \caption{The Feynman diagram of exclusive $W^+W^-$ production in photon-photon
  interaction (left) and diagrams contributing to the $\gamma\gamma\rightarrow
  W^+W^-$ subprocess via triple $\gamma WW$ (centre) and  quartic $\gamma\gamma
  WW$ (right) SM couplings.}
  \label{fig:diagWW}
\end{figure}

A particularly interesting exclusive process that could be studied at the LHC
is the production of $W$ boson pairs.  A complete review of this process can be
found in \cite{Kepka:2008yx,Chapon:2009hh}, only the main points are outlined in this note. As
mentioned above, a $WW$ pair can be produced in a photon-photon process, see
Fig.  \ref{fig:diagWW} (left). The cross section is calculated as a convolution
of the photon-photon luminosity $\mathcal{L}_{\gamma\gamma}$ and the subprocess
cross section $\hat\sigma_{\gamma\gamma\rightarrow WW}$:
\[ \sigma_{pp \rightarrow p W^+W^- p} = \int \mathcal{L}_{\gamma\gamma}
\hat\sigma_{\gamma\gamma\rightarrow WW}. \]
At tree level both the triple $\gamma WW$ (\textit{e.g.} Fig.
\ref{fig:diagWW}, centre) and the quartic  $\gamma\gamma WW$ (\textit{e.g.}
Fig.  \ref{fig:diagWW}, right) SM couplings contribute to the subprocess cross
section. This ensures the unitarity of the SM at high energies due to
cancellations between these diagrams and gives the cross section of 95.6 fb at
$\sqrt{s} = 14\ \textrm{TeV}$.

Measurements of this process will allow to test the existence of the SM quartic
$\gamma\gamma WW$ couplings, which can give an insight into the electroweak
symmetry breaking mechanism. Also, it is expected in Beyond Standard Model
(BSM) theories (especially in Higgsless and extra-dimension models)
that the cross section for exclusive $WW$ production is larger than in the
SM.

Such BSM effects can be described in terms of anomalous couplings of the gauge
bosons. This is done by adding into the SM Lagrangian new (anomalous) terms
that are responsible for the modified interactions. The cross section for the
$WW$ production in the photon-photon channel rise very quickly with the
anomalous couplings values. Currently, the best limits on these values come
from the OPAL Collaboration \cite{Abbiendi:2004bf}. With the AFP detectors one
can constrain the values of the couplings by four orders of magnitude and reach
the values predicted by Higgsless and extra-dimension models.

\section{AFP -- The Experimental Setup}

A crucial ingredient for all the studies described above is to have an
experimental possibility to detect forward protons.  Such protons are scattered
at a very small angles, of the order of microradians, and into the accelerator
beampipe. Therefore, to measure the protons one needs detectors that are placed
inside the machine beampipe. Such measurements are quite difficult but it has been
shown at HERA, Tevatron and RHIC that they are possible
\cite{Aaltonen:2007hs,Wolf:2009jm,Bultmann:2006tt,Bueltmann:2003gq}.  Although
the ATLAS Experiment already has such detectors (these are the ALFA stations
\cite{ALFA_TDR}), their purpose is to measure the elastic scattering process
and they can work only during special, very low luminosity LHC runs with a
dedicated, high $\beta^*$, machine tune. AFP is an upgrade project for the
ATLAS Experiment aiming to install additional detectors that will be able to
detect forward protons during the normal LHC runs. The primary goal for the new
detectors is to study the Central Exclusive Production, especially the
exclusive $WW$ production. 

\begin{figure}[t] \centering \includegraphics[width=\textwidth]{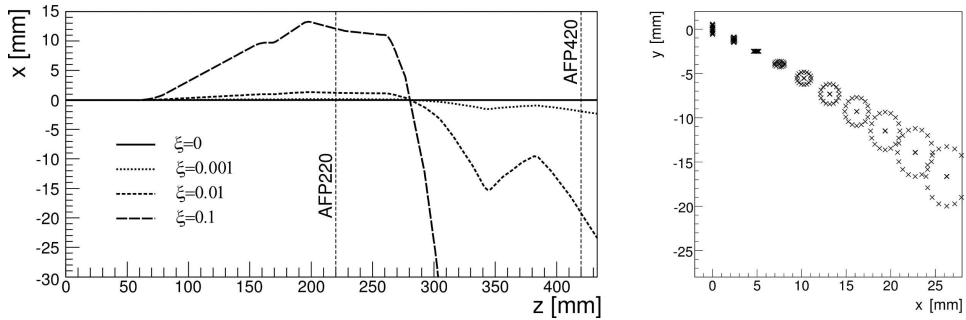}
  \caption{Left: Simulated horizontal trajectory of a 7 TeV proton and protons
  with tree different values of reduces energy loss. Right: The position in
  AFP220 of scattered protons. Moving from left to right, different ellipses
  correspond to increasing values of $\xi$, the centers of ellipses correspond
  to $t=0.0$ GeV$^2$, while the ellipses correspond to $t=0.5$ GeV$^2$.}
  \label{fig:transport} \end{figure}

A forward proton, which is scattered into the beampipe, goes through the LHC magnets
together with the proton beam. However, due to the energy loss in the
interaction, its momentum is smaller than the nominal one. Therefore, in the
magnetic fields of the LHC magnets its trajectory is bent more than that of the
protons of the beam. The transverse distance between the beam and the forward
proton increases with the distance from the Interaction Point (IP), see Fig.
\ref{fig:transport}, and at some point the forward proton hits the beampipe.
Before this happens the proton is separated enough from the beam to be tagged by
dedicated detectors. 

  The AFP \cite{AFP} project assumes installation of eight stations with such detectors
around the ATLAS IP. The installation is planned in two phases:

\begin{enumerate}
    
  \item AFP220 -- stations at 216 and 224 m (on both sides of the IP),
  
  \item AFP420 -- stations at 416 and 424 m (on both sides of the IP).

\end{enumerate}

The first phase of the installation (AFP220) is planned for the long LHC shutdown
starting in 2012. These detectors can be used for the studies of exclusive
production of jets and anomalous couplings. The second phase (AFP420) is more
complicated from the technical point of view, because 420 m is already the LHC cold
region and to install the detectors one needs to interfere with the
machine liquid helium system.

The AFP420 stations are needed for the low mass Higgs studies, but are not
necessary for the anomalous couplings study. Therefore, the second phase of the
AFP project will be possibly performed only if the SM Higgs is discovered and
the uncertainty for its production in the exclusive mode is more constrained.
The second phase could be considered also in the case of supersymmetry
discovery.

\begin{figure}[t]
  \centering
  \includegraphics[width=\textwidth]{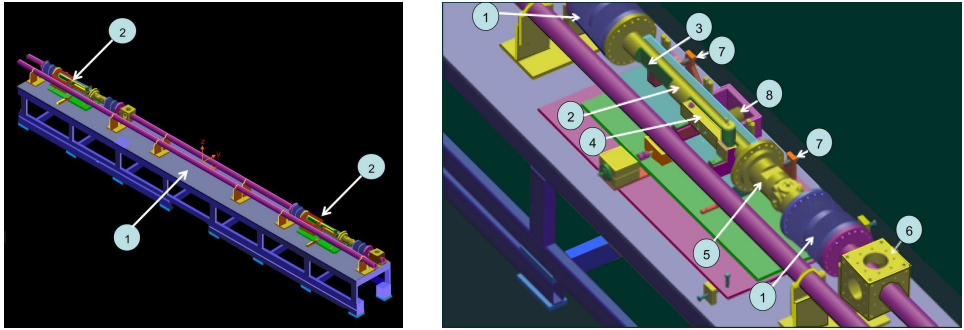}
  \caption{Left -- schematic view of the AFP220 setup: (1) detector arm with
  support table, (2) detector sections at 216 and 224 metres. Right -- top view
  of one  detector section: (1) bellows, (2) moving pipe, (3) Si-detector
  pocket, (4) timing detector, (5) moving BPM, (6) fixed BPM, (7) LVDT position
  measurement system, (8) emergency spring system.}
  \label{fig:HamburgBeampipe}
\end{figure}

To perform the measurement of forward protons one needs detectors that are
placed inside the accelerator beampipe. The smaller the distance between the
active detector area and the beam the better the acceptance that can be
obtained.  However, one must realise that in the beginning of most runs the
beam is ``hot'', which means that it can be unstable. To be able to perform the
measurements the detectors need to be movable -- it must be possible to adjust
their position (\textit{i.e.} their distance from the beam) appropriately to the
beam conditions. The AFP project assumes the use of the Hamburg movable beampipe
mechanism to adjust the horizontal position of the detectors, see Fig.
\ref{fig:HamburgBeampipe}.

For the measurement of the proton position the required resolution is 10~$\mu$m in the
horizontal and 30 $\mu$m in vertical direction. Additional important
requirements for the detectors are: high efficiency, small dead space at the
edge of the sensors (active edge) and sufficient radiation hardness. A silicon
tracking detector fulfilling these requirements will be placed in each AFP
station. Such detector will consist of five layers of active detectors that
will together give the required resolution. 

At the LHC one needs to remember about the pile-up. Not only will the bunch
crossings  occur every 25 ns, but also in each of them there will be many
independent $pp$ interactions. For studies of the Central Exclusive Production
it must be possible to tell if both protons observed in the AFP detectors come
from the same interaction (the same vertex) or it is just a random coincidence
of two intact protons coming from two different, independent $pp$ collisions. For
this purpose additional detectors are needed -- fast timing detectors with
resolution of a few picoseconds. With such detectors it is possible to measure
the proton time of flight from the vertex to the AFP station so precisely that
it will be possible to reconstruct the longitudinal position of the interaction
vertex with resolution of a few millimeters. Such measurement performed for
both protons will help to distinguish the signal from the pile-up background.

\begin{figure}[t]
  \centering
  \includegraphics[width=\textwidth]{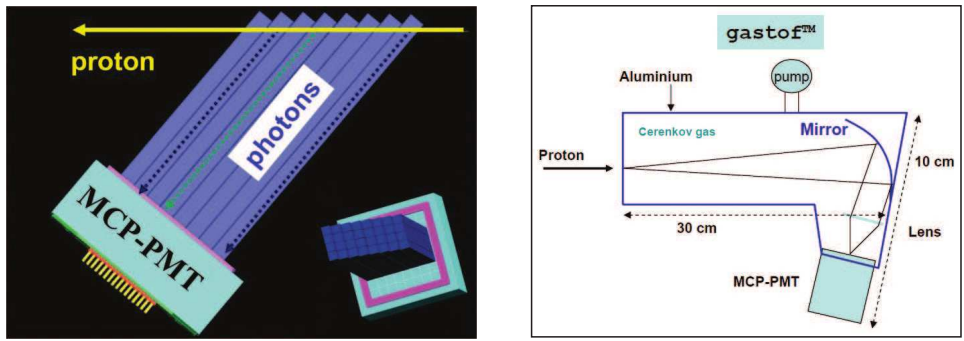}
  \caption{Left: A schematic side view of the proposed QUARTIC time-of-flight
  counter, which shows Cerenkov photons being emitted and channeled to the
  MCP-PMT as the proton traverses the eight fused silica bars in one row. The
  inset shows a rotated view with all four rows visible.  Right: A schematic
  view of the proposed GASTOF time-of-flight counter.}
  \label{fig:QuarticGastof}
\end{figure}

Two different designs of timing detectors are considered. The first one,
QUARTIC (see Fig. \ref{fig:QuarticGastof} left), consists of a matrix of quartz
bars. A proton traversing the detector creates Cherenkov light that propagates
along the bars to the photomultiplier. In the second detector, GASTOF, a gas is
used as the active material and the created Cherenkov light is focused by a
mirror and then detected by a photomultiplier (see Fig. \ref{fig:QuarticGastof}
right).

Kinematics of a forward proton is often described by means of the reduced
energy loss $\xi$:
\[ \xi = \frac{\Delta E}{E_0} = \frac{E_0 - E}{E_0},\]
where $E_0$ is the nominal energy of the beam, $E$ is the energy after the
interaction and $\Delta E$ is the energy that proton lost in the process.  For the
Central Exclusive Production there is a simple approximate relation between the
reduced energy losses of both protons ($\xi_1$ and $\xi_2$) and the mass $M$ of
the centrally produces particle/system (missing mass): 
\[ M^2 = s \xi_1 \xi_2, \]
where $s=(2E_0)^2$ is the centre of mass energy squared.

\begin{figure}[t]
  \centering
  \includegraphics[height=4.5cm]{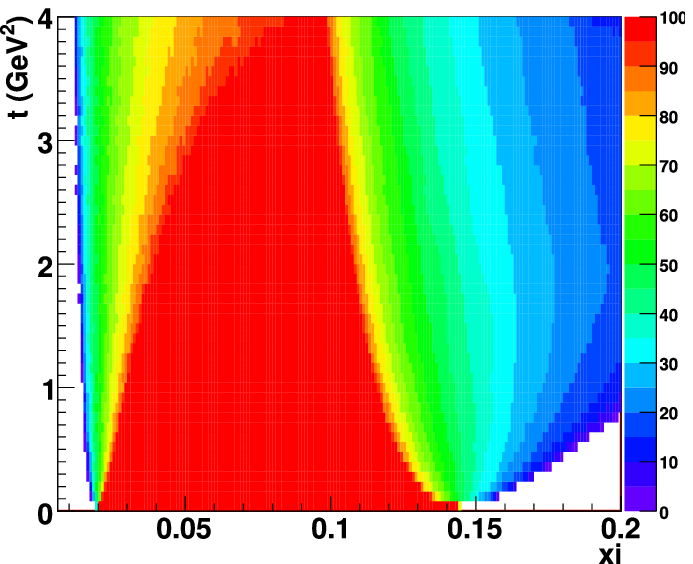}
  \hfill
  \includegraphics[height=4.5cm]{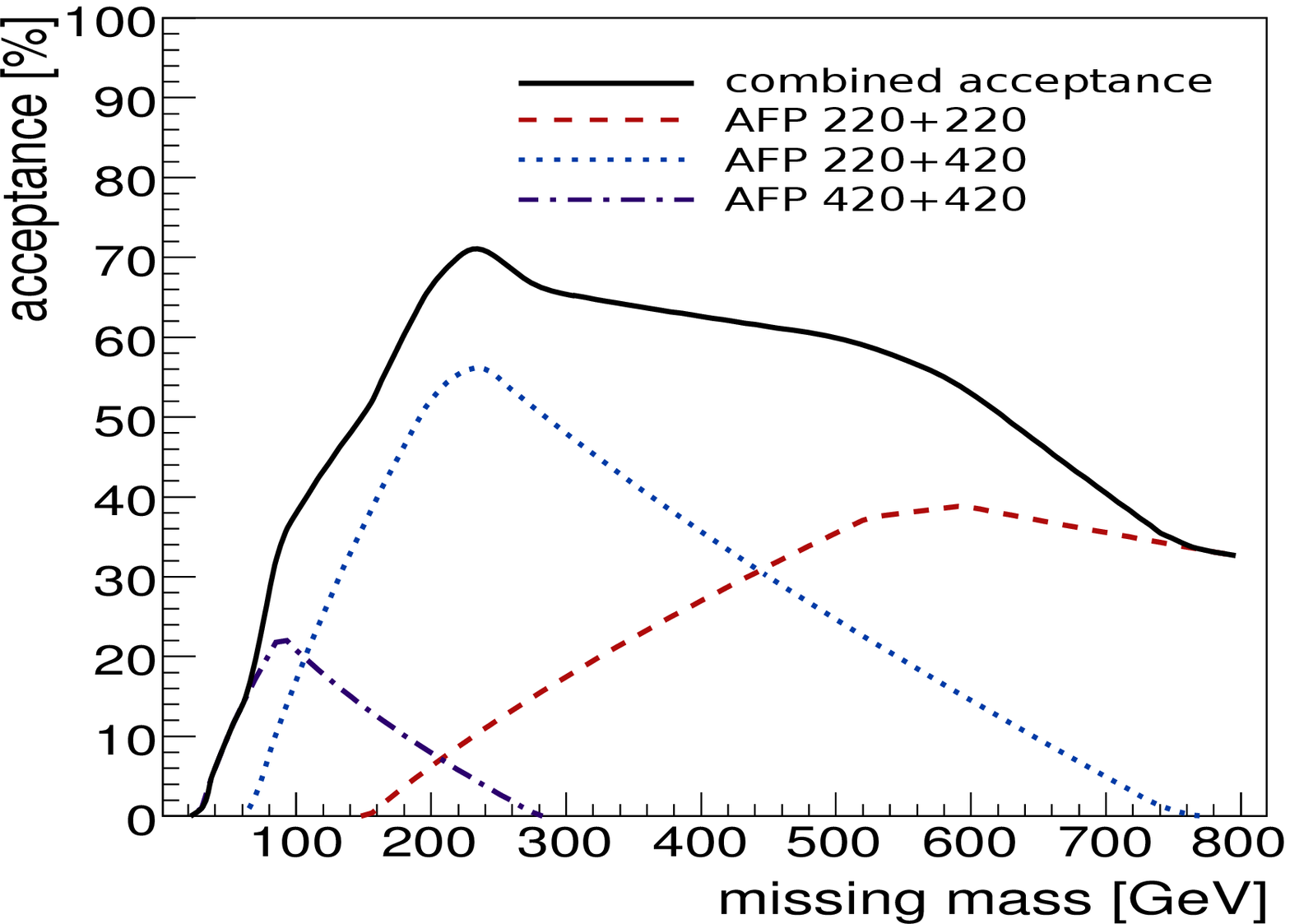}
  \caption{Left: Acceptance of the AFP220 stations on beam 1 as a function of
  $\xi$ and $t\approx p_T^2$. Right: Acceptance of the AFP detectors as a
  function of the missing mass.}
  \label{fig:accepptance}
\end{figure}

Both sets of detector stations (AFP220 and AFP420) are needed to cover a wide
range of  $\xi$. The AFP220 stations can detect protons with $\xi$
between 0.02 and 0.2, whereas the AFP420 stations accept $0.002<\xi<0.02$.
The lower endpoints of these intervals depend on the distance
between the detector and the beam (the smaller the distance, the smaller the
minimal $\xi$). Fig.  \ref{fig:accepptance} (left) presents the acceptance of
the AFP220. For the Central Exclusive production the acceptances of both
protons can be translated into the acceptance as a function of the centrally
produced mass, see Fig \ref{fig:accepptance} (right).  Three different
signatures are considered in this plot:

\begin{enumerate}
  \item AFP220+AFP220 -- both protons are detected by AFP220 stations,
  \item AFP420+AFP420 -- both protons are detected by AFP420 stations.  
  \item AFP220+AFP420 -- one proton is detected by the AFP220 station and the
    second one by the AFP420 station,
\end{enumerate}

\section{Summary}

Central Exclusive Production and photon-photon interactions in $pp$
collisions are very interesting processes to study at the LHC. Uncertainties of
the theoretical calculations are quite large, thus measurements of
\textit{e.g.} exclusive jets could lead to better understanding of the strong
interactions ruling these processes. Other possible results are determination
of the Higgs boson quantum numbers and probing new physics via the gauge bosons
anomalous couplings.

For all these to happen, new dedicated detectors are needed to tag the forward
protons. The AFP project of such an upgrade of the ATLAS experiment has been
presented. In the first phase it postulates installation of four stations on
both sides of the central ATLAS detector. Each station will consist of a
silicon detector measuring the proton position and the timing detector that
measures TOF of the scattered proton with picoseconds precision.  The detectors
will be placed in the Hamburg movable beampipe.

At the time of writing this note the Technical Proposal of the AFP was being
reviewed by the ATLAS Collaboration. If the project is accepted, the AFP220
stations will be installed in the tunnel during the next long LHC shutdown.


\begin{thebibliography}{99}

%\cite{d'Enterria:2007dt}
\bibitem{d'Enterria:2007dt}
  D.~G.~d'Enterria,
  ``\textit{Forward Physics at the LHC}''
    [arXiv:0708.0551 [hep-ex]].


\bibitem{Aaltonen:2007hs}
  CDF Collaboration,
  ``\textit{Observation of Exclusive Dijet Production at the Fermilab Tevatron p-pbar Collider}''
  Phys.\ Rev.\  {\bf D77 } (2008)  052004.

\bibitem{Khoze:2000cy}
  V.~A.~Khoze, A.~D.~Martin and M.~G.~Ryskin, ``\textit{Can the Higgs be seen
  in rapidity gap events at the Tevatron or the LHC?}'', Eur.\ Phys.\ J.\  C
  {\bf 14} (2000) 525, ``\textit{Prospects for new physics observations in
  diffractive processes at the LHC and Tevatron}'' Eur.\ Phys.\ J.\  C {\bf 23}
  (2002) 311.

\bibitem{Staszewski:2009sw}
  R.~Staszewski and J.~Chwastowski,
  ``\textit{Transport Simulation and Diffractive Event Reconstruction at the LHC}'',
  Nucl.\ Instrum.\ Meth.\  A {\bf 609}, 136 (2009)
  [arXiv:0906.2868 [physics.ins-det]].
  %%CITATION = NUIMA,A609,136;%%


\bibitem{Dechambre:2011py}
  A.~Dechambre, O.~Kepka, C.~Royon, R.~Staszewski,
  ``\textit{Uncertainties on exclusive diffractive Higgs and jets production at the LHC}'',
  Phys.\ Rev.\  {\bf D83 } (2011)  054013.
  %[arXiv:1101.1439 [hep-ph]].

\bibitem{Krasny:2006xg}
  M.~W.~Krasny, J.~Chwastowski, K.~Slowikowski,
  ``\textit{Luminosity measurement method for LHC: The Theoretical precision and the experimental challenges}'',
  Nucl.\ Instrum.\ Meth.\  {\bf A584 } (2008)  42-52.

\bibitem{Schul:2008sr}
  N.~Schul, K.~Piotrzkowski,
  ``\textit{Detection of two-photon exclusive production of supersymmetric pairs at the LHC}'',
  Nucl.\ Phys.\ Proc.\ Suppl.\  {\bf 179-180 } (2008)  289-297.

\bibitem{Kepka:2008yx}
  O.~Kepka, C.~Royon,
  ``\textit{Anomalous WW gamma coupling in photon-induced processes using forward detectors at the LHC}'',
  Phys.\ Rev.\  {\bf D78 } (2008)  073005.

  \bibitem{Chapon:2009hh}
  E.~Chapon, C.~Royon, O.~Kepka,
  ``\textit{Anomalous quartic $WW\gamma\gamma$, $ZZ\gamma\gamma$, and trilinear $WW\gamma$ couplings in two-photon processes at high luminosity at the LHC}'',
  Phys.\ Rev.\  {\bf D81 } (2010)  074003.



%\cite{Abbiendi:2004bf}
\bibitem{Abbiendi:2004bf}
  OPAL Collaboration,
  ``\textit{Constraints on anomalous quartic gauge boson couplings from nu anti-nu gamma gamma and q anti-q gamma gamma events at LEP-2}'',
  Phys.\ Rev.\  {\bf D70 } (2004)  032005.

%\cite{Wolf:2009jm}
\bibitem{Wolf:2009jm}
  G.~Wolf,
  ``\textit{Review of High Energy Diffraction in Real and Virtual Photon Proton scattering at HERA}'',
  Rept.\ Prog.\ Phys.\  {\bf 73 } (2010)  116202.

  %\cite{Bultmann:2006tt}
\bibitem{Bultmann:2006tt} %pp2pp RHIC
  pp2pp Collaboration,
  ``\textit{Double Spin Asymmetries A(NN) and A(SS) at s**(1/2) = 200-GeV in Polarized Proton-Proton Elastic Scattering at RHIC}'',
  Phys.\ Lett.\  {\bf B647 } (2007)  98-103.

  %\cite{Bueltmann:2003gq}
\bibitem{Bueltmann:2003gq}
  S.~L.~Bueltmann, I.~H.~Chiang, R.~E.~Chrien, A.~Drees, R.~L.~Gill, W.~Guryn, D.~Lynn, C.~Pearson {\it et al.},
  ``\textit{First measurement of proton proton elastic scattering at RHIC}'',
  Phys.\ Lett.\  {\bf B579 } (2004)  245-250.


\bibitem{ALFA_TDR}
ATLAS Luminosity and Forward Physics Community: \emph{ATLAS TDR 018}, CERN/LHCC/2008-004.

\bibitem{AFP}
AFP Collaboration,
``\textit{Letter of Intent for ATLAS FP: A
Project to Install Forward Proton Detectors at 220 m and 420 m Upstream and Downstream
of the ATLAS Detector}'', http://jenni.web.cern.ch/jenni/AFP.loi atlas.pdf.


\end{thebibliography}
\end{document}